\documentclass[twocolumn,preprintnumbers,amsmath,aps]{revtex4}
\usepackage{graphicx}
\usepackage{dcolumn}
\usepackage{bm}
\usepackage{color}
\usepackage{multirow}
\begin{document}
\def\eq#1{(\ref{#1})}
\def\fig#1{Fig.\hspace{1mm}\ref{#1}}
\def\tab#1{Tab.\hspace{1mm}\ref{#1}}
\title{Influence of strong-coupling and retardation effects\\ on superconducting state in ${\rm YB_{6}}$ compound}
\author{A. M. Bujak$^{\left(1\right)}$}
\author{K. A. Szewczyk$^{\left(2\right)}$}
\author{M. Kostrzewa$^{\left(2\right)}$}
\email{malgorzata.kostrzewa@ajd.czest.pl}
\author{K. M. Szcz{\c{e}}{\'s}niak$^{\left(3\right)}$} 
\author{M. A. Sowi{\'n}ska$^{\left(1\right)}$}
\affiliation{$^1$ Institute of Physics, Cz{\c{e}}stochowa University of Technology, Ave. Armii Krajowej 19, 42-200 Cz{\c{e}}stochowa, Poland}
\affiliation{$^2$ Institute of Physics, Jan D{\l}ugosz University in Cz{\c{e}}stochowa, Ave. Armii Krajowej 13/15, 42-200 Cz{\c{e}}stochowa, Poland}
\affiliation{$^3$ Ul. Pomorska 37/55, Zawiercie, 42-400, Poland}
\date{\today} 

\begin{abstract}
In the framework of strong-coupling formalism, we have calculated the thermodynamic parameters of superconducting state in 
the ${\rm YB_{6}}$ compound. The values of critical temperature ($T_{C}$) are $9.5$~K and $7.9$~K for the Coulomb pseudopotential 
$\mu^{\star}=0.1$ and $0.2$, respectively. In the paper, we have determined the low temperature values of order parameter ($\Delta(0)$), specific heat jump at the critical temperature ($\Delta C(T_{C})$), and thermodynamic critical field ($H_C(0)$). The dimensionless thermodynamic ratios: 
$R_{\Delta}=2\Delta\left(0\right)/{k_BT_C}$, $R_C=\Delta C\left(T_C\right)/C^N\left(T_C\right)$, and
$R_H=T_CC^N\left(T_C\right)/H_C^2\left(0\right)$ are equal to: 
$R_{\Delta}\left(\mu^{\star}\right)\in\lbrace 4.48,4.35\rbrace$, $R_{C}\left(\mu^{\star}\right)\in\lbrace 2.62,2.55\rbrace$, and 
$R_{H}\left(\mu^{\star}\right)\in\lbrace 0.146,0.157\rbrace$. Due to the significant strong-coupling and retardation effects 
($k_{B}T_{C}\slash \omega_{\rm ln}\sim 0.1$) those values highly deviate from the predictions of BCS theory.\\

\end{abstract}

\maketitle

\section{Introduction}

The discovery of superconductivity in the ${\rm MgB_{2}}$ compound (${\rm T_{C}\sim39}$~K) \cite{Nagamatsu2001A} contributed to the increasing 
interest on other borides. The studies conducted over the group of general formula ${\rm MB_{6}}$, where the symbol M denotes one of metals - Y, La, Th or Nd \cite{Buzea2001A, Szabo2013A}, allowed to claim that the yttrium boride reaches the second in value of critical temperature (${\rm T_{C}\sim 8.4}$~K) \cite{Gabani2014A}. It should be emphasized that ${\rm YB_{6}}$ belongs to the family of type II superconductors. The yttrium hexaboride crystallizes in the body-centered-cubic ${\rm CaB_{6}}$-type structure with the space group of $Pm3m$, where the ${\rm B_{6}}$ molecule has the octahedral structure \cite{Gabani2014A}. The distances between the atoms of yttrium and boron are relatively large ($3.01$~${\rm \AA}$) \cite{Lortz2005A, Wang2005A}. 

Historically, the first experimental result for ${\rm YB_{6}}$ comes from 1968, whereas $T_{C}\sim 7.1$~K \cite{Matthias1968A}. 
It was fully confirmed by Lortz {\it et al.} \cite{Lortz2006A}, where the single crystals of high quality were used. 
In particular, the critical temperature was determined with the help of the four different methods: 
the resistivity in zero field ($T_{C}\sim 7.2$~K), 
the AC susceptibility at $8$ kHz ($T_{C}\sim 7.24$~K), 
the Meissner magnetization at $2.7$ Oe ($T_{C}\sim 7.13$~K), 
and the specific heat jump in zero field ($T_{C}\sim 7.15$~K). 
The average value of critical temperature for the four presented series is equal to $\sim 7.2$~K. 
It is worth mentioning that Lortz {\it et al.} also studied two other crystals, obtaining $T_{C}$ from $6.5$~K to $7.6$~K, 
dependent on boron concentration. The higher value of critical temperature corresponds the lesser boron concentration, which has been confirmed by the literature data \cite{Bergman1990A}. In turn, in the paper \cite{Kadono2007A} the single crystal and $\mu SR$ method were taken under consideration. 
The observed critical temperature in the determination of magnetic susceptibility and upper critical field was $\sim 6.95$~K. So far, the highest value of $T_{C}\sim 8.4$~K was obtained by the splat-cooling of the arc-melted samples with the nominal composition \cite{Fisk1976A}. 

In order to understand the mechanism of induction of the superconducting state in yttrium boride and the possibilities of its potential uses, the lot of theoretical and experimental studies were conducted \cite{Szabo2013A, Lortz2006A, Teyssier2006A, Jager2004A, Kadono2007A, Xu2007A, Khasanov2006A}. 
The results allowed to draw the conclusion that the soft phonon modes of value equal to $7.5$ meV, which come from yttrium atoms closed in the octahedral system of boron atoms make the biggest contribution to the electron-phonon coupling \cite{Lortz2006A, Teyssier2006A, Xu2007A}. 
However, the results included in the theoretical paper \cite{Schell1982A} suggest that the main contribution to the electron-phonon coupling constant should be connected with the boron vibrations. For comparison, in ${\rm ZrB_{12}}$ compound those interactions posses the energies of the order $15$~meV \cite{Lortz2005A}, in ${\rm MgB_{2}}$ of the order of $60$~meV \cite{Geerk2005A}.

In the theoretical paper \cite{Xu2007A}, the values of $T_{C}$ at zero pressure were obtained in the range from $6.2$~K to $8.9$~K, 
which fairly enough corresponds with the experimental data \cite{Matthias1968A, Lortz2006A}. Also, the evident negative effect of pressure on the superconducting properties was found. For example, the increase of pressure in the range from $0$ to $40$~GPa causes the fall of electron-phonon coupling constant from $1.44$ to $0.44$, which corresponds to the decrease of the critical temperature from $8.9$~K to $1.9$~K. The theoretical results were 
partly confirmed experimentally by Khasanov {\it et al.} \cite{Khasanov2006A}, where the fall of critical temperature from $6.6$~K to $6.1$~K for the pressure from $0$ to $0.92$~GPa was observed. The influence of pressure on the superconducting properties of two single samples of the yttrium boride ($T_{C}\sim 7.5$~K and $T_{C}\sim 5.9$~K), created in the atmosphere of argon at the pressure of $10^{-4}$~GPa, was studied in \cite{Gabani2014A}. The obtained results coincide with the data observed by Lortz {\it et al.} \cite{Lortz2006A}.

In the presented work, we examined the most important thermodynamic properties of the superconducting state in ${\rm YB_{6}}$ compound. To our knowledge, such an analysis has never been presented in the literature on the subject. The obtained results turned out to be interesting due to the fact that despite the low value of $T_{C}$ the other thermodynamic parameters strongly differ from the predictions of the classic BCS theory 
\cite{Bardeen1957A, Bardeen1957B}.

\section{Formalism}

Superconducting state in the ${\rm YB_{6}}$ compound is induced by the strong electron-phonon interaction. 
In particular, the electron-phonon coupling constant $\lambda$ equals $1.4$, where ${\lambda=2\int^{+\infty}_0\alpha^2F\left(\omega\right)d\omega/\omega}$. The symbol $\alpha^{2}F\left(\omega\right)$ denotes the Eliashberg function, which was determined by Xu {\it et al.} \cite{Xu2007A}, with the use of pseudopotential plane-wave method (Quantum-Espresso package) \cite{Hohenberg1964A, Baroni1987A, Giannozzi1991A}. From the physical side, the Eliashberg function models the linear electron-phonon interaction in a quantitative way. However, it does not contain any information about the electron correlations occurring in the studied system.
 
Due to the high value of the electron-phonon coupling constant, the properties of the superconducting state in ${\rm YB_{6}}$ were characterized in the framework of the Eliashberg equations defined on the imaginary axis and in the mixed representation  
\cite{Carbotte1990A, Eliashberg1960A, Marsiglio1988A}. On the imaginary axis, the Eliashberg equations take the following form: 
\begin{equation}
\label{r1}
\varphi_{n}=\frac{\pi}{\beta}\sum_{m=-M}^{M}
\frac{K\left(i\omega_{n}-i\omega_{m}\right)-\mu^{\star}\theta\left(\omega_{c}-|\omega_{m}|\right)}
{\sqrt{\omega_m^2Z^{2}_{m}+\varphi^{2}_{m}}}\varphi_{m},
\end{equation}
and
\begin{equation}
\label{r2}
Z_{n}=1+\frac{1}{\omega_{n}}\frac{\pi}{\beta}\sum_{m=-M}^{M}
\frac{K\left(i\omega_{n}-i\omega_{m}\right)}{\sqrt{\omega_m^2Z^{2}_{m}+\varphi^{2}_{m}}}
\omega_{m}Z_{m}.
\end{equation}
The order parameter is defined by the ratio: $\Delta_{n}=\varphi_{n}/Z_{n}$, where 
$\varphi_{n}=\varphi\left(i\omega_{n}\right)$ denotes the order parameter's function, and $Z_{n}=Z\left(i\omega_{n}\right)$ is the wave function renormalization factor. The symbol $\omega_{n}$ is the fermion Matsubara frequency: $\omega_{n}=\pi/\beta\left(2n-1\right)$, 
while $\beta=\left(k_{B}T\right)^{-1}$, and $k_{B}$ is the Boltzmann constant. In contrast to the standard BCS model, the order parameter in the Eliashberg formalism explicitly depends on the energy, which means that the theory considers the retardation of the pairing interaction. In addition, the renormalization of the value of $\varphi_{n}$ by $Z_{n}$ is the typical strong-coupling effect: $Z_{n=1}\simeq 1+\lambda$ \cite{Carbotte1990A}. 

The electron-phonon interaction determines directly the values of pairing kernel:
$K\left(z\right)=2\int_0^{+\infty}d\omega\frac{\omega}{\omega ^2-z^{2}}\alpha^{2}F\left(\omega\right)$. It is worth noting that the form of the pairing kernel is not universal. In the case considered by us, we assume that the function $K\left(z\right)$ was derived in the approximation 
of non-interacting phonons. This means that in the considerations we omit the self-energy of the phonon propagator.

We have taken into account two values of Coulomb pseudopotential: $\mu^{\star}=0.1$ and $0.2$. From the physical side, $\mu^{\star}$ models the depairing electron correlations in the easiest way possible (i.e. parametric) \cite{Morel1962A}. $\theta$ denotes the Heaviside function. The cut-off frequency is equal to $\omega_{c}=5\Omega_{\rm{max}}$, where $\Omega_{\rm{max}}$ is the maximum phonon frequency. The choice of the cut-off frequency is to some extent arbitrary. By default, it is assumed that $\omega_{c}\in\left(3\Omega_{\rm{max}},10\Omega_{\rm{max}}\right)$ \cite{Carbotte1990A}. Hence, the critical temperature value obtained should always be given together with the value of $\mu^{\star}$ and $\omega_{c}$.

The Eliashberg equations on the imaginary axis have been solved for $M=1100$ Matsubara frequencies in the range of temperature 
from $T_{0}=2.25$~K to $T_{C}$. Of course, going down below the temperature $T_{0}$ is technically possible. However, it would require the significant increase in the value of $M$, which is not necessary from the physical point of view, because in the low temperature range, the order parameter becomes saturated. Performing calculations we have used the numerical methods tested in the papers  \cite{Szczesniak2001A, Szczesniak2011B, Szczesniak2012I, Durajski2012A, Szczesniak2012N}.

In the imaginary axis formalism, it can be assumed with the good approximation that the maximum value of the order parameter $\Delta_{n=1}$ corresponds to the physical value of the order parameter. In contrast, the maximum value of the wave function renormalization factor 
($Z_{n=1}$) determines the ratio of the effective mass of the electron ($m^{\star}_{e}$) to the electron band mass ($m_{e}$). 
The above quantities can be very precisely calculated in the framework of the real axis formalism. In the presented work, the transition to the real axis was carried out using the Eliashberg equations in the mixed representation \cite{Marsiglio1988A}:
\begin{eqnarray}
\label{r3}
& &\varphi\left(\omega\right)=\\ \nonumber
                                 & &\frac{\pi}{\beta}\sum_{m=-M}^{M}
                                  \left[K\left(\omega-i\omega_{m}\right)-\mu^{\star}\left(\omega_{m}\right)\right]
                                  \frac{\varphi_{m}}
                                  {\sqrt{\omega_m^2 Z^{2}_{m}+\varphi^{2}_{m}}}\\ \nonumber
&+&i\pi\int_{0}^{+\infty}d\omega^{'}\alpha^{2}F\left(\omega^{'}\right)
                                \Bigg[\left[f_{\rm BE}\left(\omega^{'}\right)+f_{\rm FD}\left(\omega^{'}-\omega\right)\right]\\\nonumber
                                &\times&\frac{\varphi\left(\omega-\omega^{'}\right)}
                                  {\sqrt{\left(\omega-\omega^{'}\right)^{2}Z^{2}\left(\omega-\omega^{'}\right)
                                  -\varphi^{2}\left(\omega-\omega^{'}\right)}}\Bigg]\\ \nonumber
                              &+&i\pi\int_{0}^{+\infty}d\omega^{'}\alpha^{2}F\left(\omega^{'}\right)
                                  \Bigg[\left[f_{\rm BE}\left(\omega^{'}\right)+f_{\rm FD}\left(\omega^{'}+\omega\right)\right]\\\nonumber
                                  &\times&\frac{\varphi\left(\omega+\omega^{'}\right)}
                                  {\sqrt{\left(\omega+\omega^{'}\right)^{2}Z^{2}\left(\omega+\omega^{'}\right)
                                  -\varphi^{2}\left(\omega+\omega^{'}\right)}}\Bigg],
\end{eqnarray}
and
\begin{eqnarray}
\label{r4}
& &Z\left(\omega\right)=
                                  1\\ \nonumber
                                  &+&\frac{i}{\omega}\frac{\pi}{\beta}\sum_{m=-M}^{M}
                                  K\left(\omega-i\omega_{m}\right)
                                  \frac{\omega_{m}Z_{m}}
                                  {\sqrt{\omega_m^2Z^{2}_{m}+\varphi^{2}_{m}}}\\ \nonumber
                              &+&\frac{i\pi}{\omega}\int_{0}^{+\infty}d\omega^{'}\alpha^{2}F\left(\omega^{'}\right)
                                  \Bigg[\left[f_{\rm BE}\left(\omega^{'}\right)+f_{\rm FD}\left(\omega^{'}-\omega\right)\right] \\\nonumber
                                 &\times&\frac{\left(\omega-\omega^{'}\right)Z\left(\omega-\omega^{'}\right)}
                                  {\sqrt{\left(\omega-\omega^{'}\right)^{2}Z^{2}\left(\omega-\omega^{'}\right)
                                  -\varphi^{2}\left(\omega-\omega^{'}\right)}}\Bigg]\\ \nonumber
                              &+&\frac{i\pi}{\omega}\int_{0}^{+\infty}d\omega^{'}\alpha^{2}F\left(\omega^{'}\right)
                                  \Bigg[\left[f_{\rm BE}\left(\omega^{'}\right)+f_{\rm FD}\left(\omega^{'}+\omega\right)\right]\\\nonumber
                              &\times &\frac{\left(\omega+\omega^{'}\right)Z\left(\omega+\omega^{'}\right)}
                                  {\sqrt{\left(\omega+\omega^{'}\right)^{2}Z^{2}\left(\omega+\omega^{'}\right)
                                  -\varphi^{2}\left(\omega+\omega^{'}\right)}}\Bigg].
\end{eqnarray}
The symbols $f_{BE}\left(\omega\right)$ and $f_{FD}\left(\omega\right)$ are the Bose-Einstein and Fermi-Dirac functions, respectively. 
The input parameters in equations \eq{r3} and \eq{r4} are also $\varphi_{n}$ and $Z_{n}$. 

\section{The results}

\begin{figure} 
\includegraphics[width=\columnwidth]{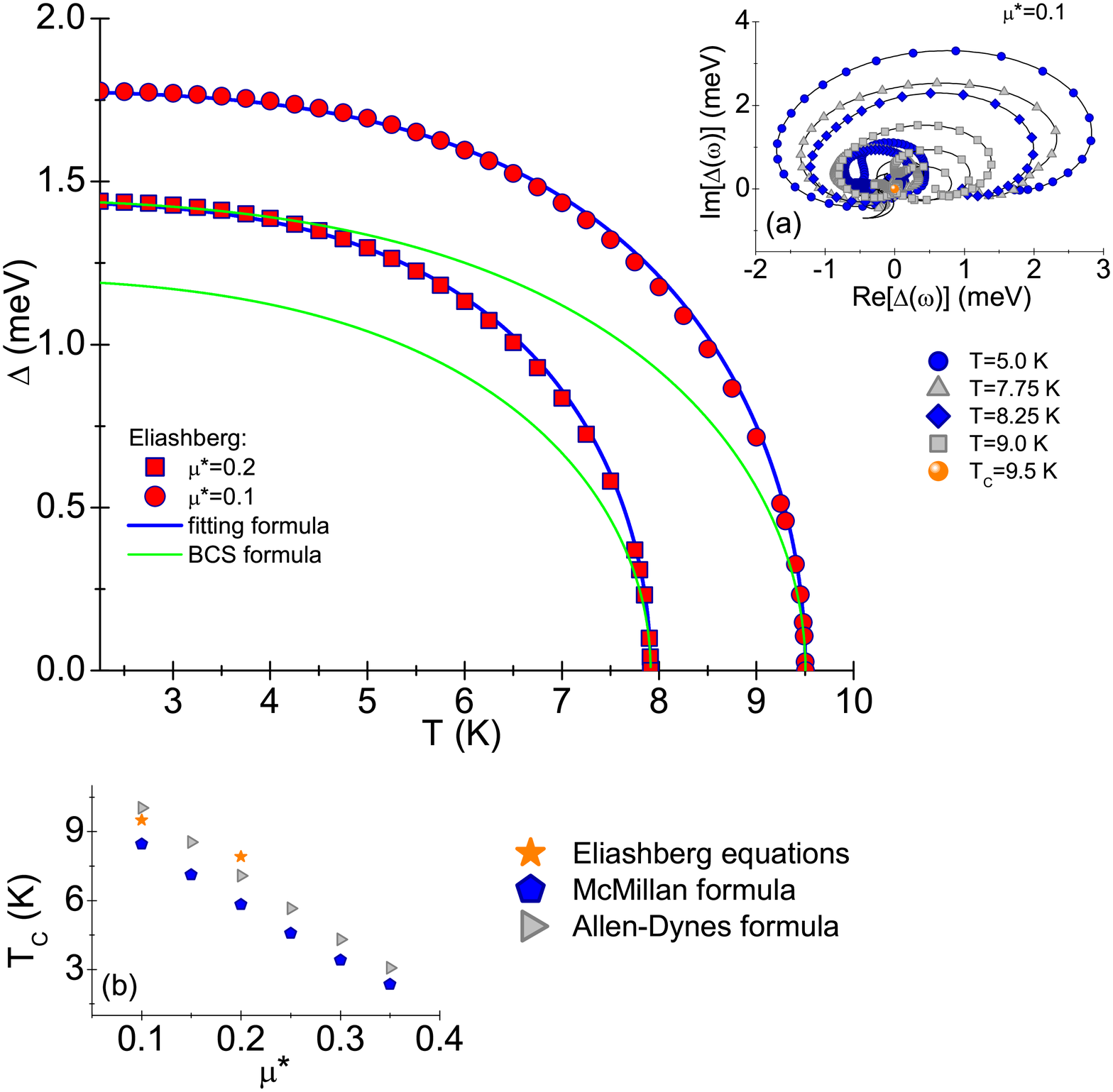}
\caption{\label{f1}
                   The influence of temperature on the order parameter. The numerical data can be parameterized with the help of 
                   the following formula: $\Delta\left(T,\mu^{\star}\right)=\Delta\left(\mu^{\star}\right)\sqrt{1-\left(T/T_{C}\right)^{\Gamma}}$, 
                   where $\Delta\left(\mu^{\star}\right)=-3.4\mu^{\star}+2.1$ and $\Gamma=3.6$ (blue lines). The BCS theory predicts $\Gamma=3$ 
                   \cite{Eschrig2001A} (green lines). 
                   Insertion (a) presents the order parameter on the complex plane. 
                   Insertion (b) the values of critical temperature calculated in the framework of full Eliashberg formalism 
                   \cite{Carbotte1990A, Eliashberg1960A}, and with the help of McMillan and Allen-Dynes formulas 
                   \cite{McMillan1968A, Allen1975A}.}
\end{figure}

\fig{f1} presents the plot of temperature courses of the order parameter. The physical values of the order parameter were calculated using the equation: $\Delta\left(T\right)={\rm Re}\left[\Delta\left(\omega=\Delta\left(T\right)\right)\right]$ \cite{Carbotte1990A}. 
In \fig{f1}~(a) we have placed sample values of the order parameter plotted on the complex plane. 
The phase transition between the superconducting and normal state is of second order. However, the obtained curves ($\Delta\left(T\right)$) do not coincide with the results predicted by the BCS mean-field model  \cite{Eschrig2001A}. The result obtained is due to the significant strong-coupling and retardation effects, which are present in ${\rm YB_{6}}$ compound. In the framework of the Eliashberg formalism those effects can be characterized with the help of critical temperature to logarithmic frequency ratio ($r=k_{B}T_{C}/{\omega_{\rm{ln}}}$), where $\omega_{{\rm ln}}=\exp\left[\frac{2}{\lambda}\int^{+\infty}_{0}d\omega\frac{\alpha^{2}F\left(\omega\right)}{\omega}\ln\left(\omega\right)\right]=6.8$~meV \cite{Carbotte1990A}. In the considered case, we have obtained $r=0.12$ and $r=0.1$, respectively for $\mu^{\star}=0.1$ and $\mu^{\star}=0.2$. In the BCS limit it is necessary to adopt $r=0$. 
Note that, in principle, the entire thermodynamics of the superconducting state is determined by the form of the order parameter. This means that all thermodynamic parameters of the superconducting phase induced in ${\rm YB_{6}}$ will be very heavily modified by the strong-coupling and retardation effects. Below we present the detailed analysis of the problem raised. 

In the first step, we set the critical temperature values. On the basis of the condition  $\Delta\left(T_{C}\right)=0$ we have obtained: 
$\left[T_{C}\right]_{\mu^{\star}=0.1}=9.5$~K and $\left[T_{C}\right]_{\mu^{\star}=0.2}=7.9$~K. The critical temperature values can also be estimated by using the McMillan or Allen-Dynes formulas  \cite{McMillan1968A, Allen1975A} (see the insertion (b) in \fig{f1}). The relatively accurate results have been reproduced merely by the Allen-Dynes expression.

The significant strong-coupling and retardation effects in ${\rm YB_{6}}$ make the ratio of energy gap to critical temperature 
($R_{\Delta}=2\Delta(0)/k_{B}T_{C}$) highly deviate from the universal constant at $3.53$, which is predicted by the BCS theory 
\cite{Bardeen1957A, Bardeen1957B}. In our case, we have: $\left[R_{\Delta}\right]_{\mu^{\star}=0.1}=4.48$ and $\left[R_{\Delta}\right]_{\mu^{\star}=0.2}=4.35$.

\begin{figure} 
\includegraphics[width=\columnwidth]{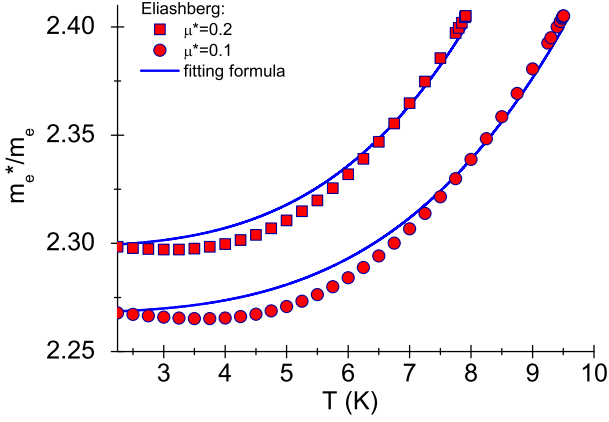}
\caption{\label{f2} 
                  The influence of temperature on the ratio of electron effective mass to electron band mass. 
                  The numerical data can be parameterized with the formula: 
                  $m_{e}^{\star}/m_{e}=\left[Z\left(T_{C}\right)-Z\left(\mu^{\star}\right)\right]\left(T/T_{C}\right)^{\Gamma}+Z\left(\mu^{\star}\right)$, 
                  where $Z\left(T_{C}\right)=1+\lambda$ and $Z\left(\mu^{\star}\right)=0.3\mu^{\star}+2.2$ (blue lines).
}
\end{figure}

In \fig{f2}, the values of the ratio of electron effective mass to electron band mass as a function of temperature have been presented
($m_{e}^{\star}\slash m_{e}={\rm Re}\left[\left[Z\left(T\right)\right]_{\omega=0}\right]$). The quantity $m_{e}^{\star}\slash m_{e}$ takes relatively 
high values due to the strong-coupling and retardation effects. It is worth to notice that, in the opposite to the order parameter, 
$m_{e}^{\star}\slash m_{e}$ weakly depends on the temperature and the Coulomb pseudopotential. 

\begin{figure} 
\includegraphics[width=\columnwidth]{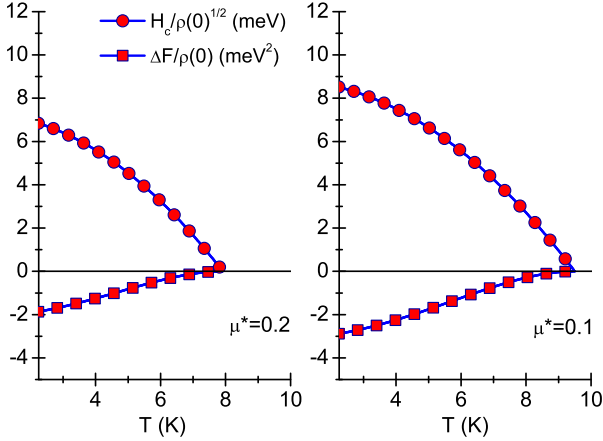}
\caption{\label{f3} The dependence of free energy difference (lower panel) and thermodynamic critical field (upper panel) on the temperature.}
\end{figure}

\begin{figure} 
\includegraphics[width=\columnwidth]{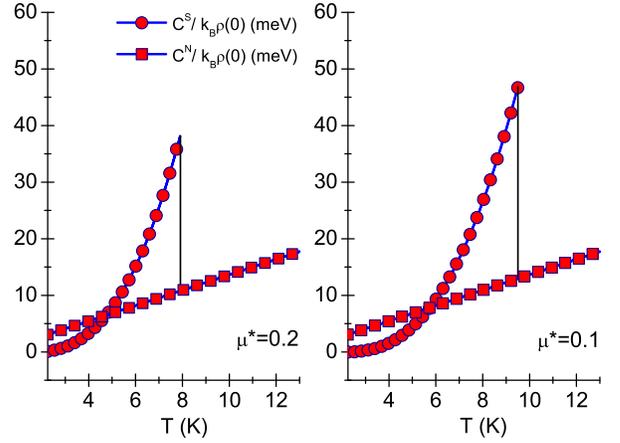}
\caption{\label{f4} The specific heat of superconducting state and normal state as a function of temperature. The vertical lines indicate 
                    the values of critical temperature.}
\end{figure}

The thermodynamic critical field and the specific heat were calculated on the basis of free energy difference between the superconducting and normal state \cite{Bardeen1964A}:
\begin{eqnarray}
\label{r5}
\Delta F/\rho\left(0\right)&=&-\frac{2\pi}{\beta}\sum^{M}_{n=1}
[\sqrt{\omega^2_n+\left(\Delta_n\right)^2}-|\omega_n|]\\ \nonumber
&\times&[Z^{\left(S\right)}_n-Z^{\left(N\right)}_n \frac{|\omega_n|}{\sqrt{\omega^2_n+\left(\Delta_n\right)^2}}],
\end{eqnarray}
where $Z^{S}_{n}$ and $Z^{N}_{n}$ denote the wave function renormalization factor for the superconducting state ($S$) and the normal state ($N$), respectively. The symbol $\rho(0)$ represents the electron density of states at the Fermi level. 

The thermodynamic critical field connects with the free energy difference in the following way:
\begin{equation}
\label{r6}
\frac{H_{C}}{\sqrt{\rho\left(0\right)}}=\sqrt{-8\pi\left[\Delta F/\rho\left(0\right)\right]}.
\end{equation}
In order to determine the specific heat difference between the superconducting state ($C^{S}$) and the normal state ($C^{N}$), it is necessary to use 
the formula:
\begin{equation}
\label{r7}
\frac{\Delta C\left(T\right)}{k_{B}\rho\left(0\right)}=-\frac{1}{\beta}\frac{d^{2}\left[\Delta F/\rho\left(0\right)\right]}{d\left(k_{B}T\right)^{2}},
\end{equation}
where $C^{N}=\gamma{T}$, and $\gamma=\frac{2}{3}\pi^{2}k_{B}^{2}\rho(0)\left(1+\lambda\right)$. 

The results have been presented in \fig{f3} and \fig{f4}. Attention has been drawn to the fact that the increase in the value of $\mu^{\star}$, which is inversely correlated with the magnitude of strong-coupling and retardation effects, causes the significant decrease in the value of free energy 
($\left[\Delta F\left(0\right)\right]_{\mu^{\star}=0.1}/\left[\Delta F\left(0\right)\right]_{\mu^{\star}=0.2}=1.55$). Of course, it translates 
in the direct way to the comparatively large drop in the value of thermodynamic critical field and specific heat jump at the critical temperature 
($\left[H_{C}\left(0\right)\right]_{\mu^{\star}=0.1}/\left[H_{C}\left(0\right)\right]_{\mu^{\star}=0.2}=1.24$ and 
 $\left[\Delta C\left(T_{C}\right)\right]_{\mu^{\star}=0.1}/\left[\Delta C\left(T_{C}\right)\right]_{\mu^{\star}=0.2}=1.24$).

The determined thermodynamic functions allowed us to calculate the dimensionless ratios: 
$R_C=\Delta C\left(T_C\right)/C^N\left(T_C\right)$ and $R_H=T_CC^N\left(T_C\right)/H_C^2\left(0\right)$. 
It turns out that the obtained results, due to the significant strong-coupling and retardation effects, very clearly differ from the predictions of BCS model ($R_{C}=1.43$ and $R_{H}=0.168$) \cite{Bardeen1957A, Bardeen1957B}. In particular, we have obtained 
$R_{C}\in\lbrace2.62,2.55\rbrace$ and $R_{H}\in\lbrace 0.146, 0.157\rbrace$, respectively for $\mu^{\star}=0.1$ and $\mu^{\star}=0.2$.

\section{Summary}

The superconducting state in $\rm YB_{6}$ compound characterizes with the relatively low value of critical temperature 
$T_{C}\sim 7.9$-$9.5$~K. Despite this, its thermodynamic properties visibly deviate from the predictions of BCS model. 
The fact above is related to the strong-coupling and retardation effects present in yttrium hexaboride ($r\sim 0.1$). 
Note that the deviations of thermodynamic parameters from the results of BCS theory can be observed in the best way by calculating the dimensionless ratios: $R_{\Delta}$, $R_C$, and $R_H$. The analysis carried out allowed to obtain: 
$R_{\Delta}\left(\mu^{\star}\right)\in\lbrace 4.48,4.35\rbrace$, $R_{C}\left(\mu^{\star}\right)\in\lbrace2.62,2.55\rbrace$, and $R_{H}\left(\mu^{\star}\right)\in\lbrace 0.146,0.157\rbrace$, in respect to the Coulomb pseudopotential equal to $0.1$ and $0.2$. On the other hand, the BCS theory predicts: $R_{\Delta}=3.53$, $R_{C}=1.43$, and $R_{H}=0.168$.


\end{document}